\documentclass[12pt]{article}
\usepackage{fullpage}
\usepackage{amsmath}
\usepackage{graphicx}
\usepackage{algorithm}
\usepackage{algorithmic}
\usepackage{rotating}

\begin{document}

\title{A Comparative Study of Collaborative Filtering  Algorithms}
\author{Joonseok Lee, Mingxuan Sun, Guy Lebanon}
\date{May 14, 2012}
\maketitle

\begin{abstract}
Collaborative filtering is a rapidly advancing research area. Every
year several new techniques are proposed and yet it is not clear
which of the techniques work best and under what conditions. In this
paper we conduct a study comparing several collaborative filtering
techniques -- both classic and recent state-of-the-art -- in a
variety of experimental contexts. Specifically, we report
conclusions controlling for number of items, number of users,
sparsity level, performance criteria, and computational complexity.
Our conclusions identify what algorithms work well and in what
conditions, and contribute to both industrial deployment
collaborative filtering algorithms and to the research community.
\end{abstract}

\section{Introduction}
\label{sec:intro}

Collaborative filtering is a rapidly advancing research area. Classic
methods include neighborhood methods such as memory based or user
based collaborative filtering. More recent methods often revolve
around matrix factorization including singular value decomposition and
non-negative matrix factorization. New methods are continually
proposed, motivated by experiments or theory. However, despite the
considerable research momentum there is no consensus or clarity on which method works best.

One difficulty that is undoubtedly central to the lack of clarity is
that the performance of different methods differ substantially based
on the problem parameters. Specifically, factors such as number of
users, number of items, and sparsity level (ratio of observed to
total ratings) affect different collaborative filtering methods in
different ways. Some methods perform better in sparse setting while
others perform better in dense settings, and so on.

Existing experimental studies in collaborative filtering either do
not compare recent state-of-the-art methods, or do not investigate
variations with respect to the above mentioned problem parameters.
In this paper we do so, and concentrate on comparing both classic
and recent state-of-the-art methods. In our experiments we control
for the number of users, the number of items, sparsity level and
consider multiple evaluation measure, and computational cost.

Based on our comparative study we conclude the following.

\begin{enumerate}
\item Generally speaking, Matrix-Factorization-based methods perform
best in terms of prediction accuracy. Nevertheless, in some special
cases, several other methods have a distinct advantage over
Matrix-Factorization methods.

\item The prediction accuracy of the different algorithms depends on the number of users, the number of items, and density, where the nature and degree of dependence differs from algorithm to algorithm.

\item There exists a complex relationship between
prediction accuracy, its variance, computation time, and memory
consumption that is crucial for choosing the most appropriate recommendation system algorithm.
\end{enumerate}

The following sections describe in detail the design and implementation of the experimental study and the experimental results themselves.

\section{Background and Related Work}
\label{sec:bg}

Before describing our experimental study, we briefly introduce
recommendation systems and collaborative filtering techniques.

\subsection{Recommendation Systems}

Broadly speaking, any software system which actively suggests an
item to purchase, to subscribe, or to invest can be regarded as a
recommender system. In this broad sense, an advertisement also
can be seen as a recommendation. We mainly consider, however, a
narrower definition of  ``personalized''
recommendation system that base recommendations on user specific
information.

There are two main approaches to personalized recommendation systems:
content-based filtering and collaborative filtering. The former makes
explicit use of domain knowledge concerning users or items. The domain knowledge may correspond to user information such
as age, gender, occupation, or location \cite{Debnath2008}, or to item
information such as genre, producer, or length in the case of movie
recommendation.

The latter category of collaborative filtering (CF) does not use user
or item information with the exception of a partially observed rating
matrix. The rating matrix holds ratings of items (columns) by users (rows) and is typically binary, for example like vs. do
not like, or ordinal, for example, one to five stars in Netflix movie
recommendation. The rating matrix may also be gathered implicitly
based on user activity, for example a web search followed by click through may be interpreted as a positive value judgement for the chosen
hyperlink \cite{Hu2008}. In general, the rating matrix is extremely
sparse, since it is unlikely that each user experienced and provided
ratings for all items.

Hybrid collaborative and content-based filtering strategies combine
the two approaches above, using both the rating matrix, and user and
item information.
\cite{Pavlov2002,Ziegler2004,Pennock2000,Melville2001,Kim2004,Basu1998,Pennock2000,Su2007}.
Such systems typically obtain improved prediction accuracy over
content-based filtering systems and over collaborative filtering
systems.

In this paper we focus on a comparative study of collaborative
filtering algorithms. There are several reason for not including
content-based filtering. First, a serious comparison of
collaborative filtering systems is a challenging task in itself.
Second, experimental results of content-based filtering are
intimately tied to the domain and are not likely to transfer from
one domain to another. Collaborative filtering methods, on the other
hand, use only the rating matrix which is similar in nature across
different domains.

\subsection{Collaborative Filtering}

Collaborative filtering systems are usually categorized into two subgroups:
memory-based and model-based methods.

Memory-based methods simply memorize the rating matrix and issue
recommendations based on the relationship between the queried user
and item and the rest of the rating matrix. Model-based methods fit
a parameterized model to the given rating matrix and then issue
recommendations based on the fitted model.

The most popular memory-based CF methods are neighborhood-based
methods, which predict ratings by referring to users whose ratings
are similar to the queried user, or to items that are similar to the
queried item. This is motivated by the assumption that if two users
have similar ratings on some items they will have similar ratings on
the remaining items. Or alternatively if two items have similar
ratings by a portion of the users, the two items will have similar
ratings by the remaining users.

Specifically, user-based CF methods \cite{Breese1998} identify users
that are similar to the queried user, and estimate the desired
rating to be the average ratings of these similar users. Similarly,
item-based CF~\cite{Sarwar2001} identify items that are similar to
the queried item and estimate the desired rating to be the average
of the ratings of these similar items.  Neighborhood methods vary
considerably in how they compute the weighted average of ratings.
Specific examples of similarity measures that influence the
averaging weights are include Pearson correlation, Vector cosine,
and Mean-Squared-Difference (MSD). Neighborhood-based methods can be
extended with default votes, inverse user frequency, and case
amplification~\cite{Breese1998}. A recent neighborhood-based method
\cite{Sun2011} constructs a kernel density estimator for incomplete
partial rankings and predicts the ratings that minimize the
posterior loss.

Model-based methods, on the other hand, fit a parametric model to the
 training data that can later be used to predict unseen ratings and
 issue recommendations. Model-based methods include
cluster-based
CF~\cite{Ungar1998,Chee2001,Connor2001,Sarwar2002,Xue2005}, Bayesian
classifiers~\cite{Miyahara2000,Miyahara2002}, and regression-based
methods \cite{Vucetic2005}. The slope-one method \cite{Lemire2005}
fits a linear model to the rating matrix, achieving fast computation
and reasonable accuracy.

A recent class of successful CF models are based on low-rank matrix
factorization. The regularized SVD method \cite{Billsus1998}
factorizes the rating matrix into a product of two low rank matrices
(user-profile and item-profile) that are used to estimate the
missing entries.  An alternative method is Non-negative Matrix
Factorization (NMF)~\cite{Lee1999} that differs in that it constrain
the low rank matrices forming the factorization to have non-negative
entries. Recent variations are Probabilistic Matrix Factorization
(PMF)~\cite{Salakhutdinov2008a}, Bayesian
PMF~\cite{Salakhutdinov2008b}, Non-linear PMF~\cite{Lawrence2009},
Maximum Margin Matrix Factorization
(MMMF)~\cite{Srebro2005,Rennie2005,DeCoste2006}, and Nonlinear
Principal Component Analysis (NPCA)~\cite{Yu2009}.

\subsection{Evaluation Measures}


The most common CF evaluation measure for prediction accuracy are
the mean absolute error (MAE) and root of the mean square error
(RMSE):
\begin{align*}
MAE &= \frac{1}{n} \sum_{u, i} |p_{u, i} - r_{u, i}|\\
RMSE &= \sqrt{\frac{1}{n} \sum_{u, i} (p_{u, i} - r_{u, i})^2}
\end{align*}
where $p_{u, i}$ and $r_{u, i}$ are the predicted and observed rating
for user $u$ and item $i$, respectively. The sum above ranges over a
labeled set that is set aside for evaluation purposes (test
set). Other evaluation measures are precision, recall, and F1 measures
~\cite{Breese1998}.

Gunawardana and Shani~\cite{Gunawardana2009} argue that different
evaluation metrics lead different conclusion concerning the relative
performance of the CF algorithms. However, most CF research papers
motivate their algorithm by examining a single evaluation measure.
In this paper we consider the performance of different CF algorithms
as a function of the problem parameters, measured using several
different evaluation criteria.

\subsection{Related Work}

Several well-written surveys on recommendation systems are
available. Adomavicius and Tuzhilin~\cite{Adomavicius2005}
categorized CF algorithms available as of 2006 into content-based,
collaborative, and hybrid and summarized possible extensions. Su and
Khoshgoftaar~\cite{Su2009} concentrated more on CF methods,
including memory-based, model-based, and hybrid methods. This survey
contains most state-of-the-art algorithms available as of 2009,
including Netflix prize competitors. A recent textbook on
recommender systems introduces traditional techniques and explores
additional issues like privacy concerns~\cite{Jannach2011}.

There are a couple of experimental studies available. The first
study by Breese et al.~\cite{Breese1998} compared two popular
memory-based methods (Pearson correlation and vector similarity) and
two classical model-based methods (clustering and Bayesian network)
on three different dataset. A more recent experimental comparison of
CF algorithms~\cite{Huang2007} compares user-based CF, item-based
CF, SVD, and several other model-based methods, focusing on
e-commerce applications. It considers precision, recall, F1-measure
and rank score as evaluation measures, with comments about the
computational complexity issue. This however ignores some standard
evaluation measures such as MAE or RMSE.

\subsection{Netflix Prize and Dataset}

The Netflix competition \footnote{http://www.netflixprize.com} was
held between October 2006 and July 2009, when BellKor's Pragmatic
Chaos team won the million-dollar-prize. The goal of this
competition was improving the prediction accuracy (in terms of RMSE)
of the Netflix movie recommendation system by 10\%. The winning team
used a hybrid method that used temporal dynamics to account for
dates in which ratings were reported ~\cite{Koren2008,Bell2010}. The
second-placed place, The Ensemble~\cite{Sill2009}, achieved
comparable performance to the winning team, by linearly combining a
large number of models.

Although the Netflix competition has finished, the dataset used in
that competition is still used as a standard dataset for  evaluating CF methods. It has 480,046 users and 17,770
items with 95,947,878 ratings. This represents a sparsiy level of
1.12\% (total number of entries divided by observed entries). Older
and smaller standard datasets include MovieLens (6,040 users, 3,500
items with 1,000,000 ratings), EachMovie (72,916 users, 1,628 items
with 2,811,983 ratings), and BookCrossing (278,858 users, 271,379
items with 1,149,780 ratings).

\section{Experimental Study}
\label{sec:exp}

We describe below some details concerning our experimental study, and
then follow with a description of our major findings.

\subsection{Experimental Design}
\label{subsec:expdesign}


To conduct our experiments and to facilitate their reproducability we
implemented the PREA\footnote{http://prea.gatech.edu} toolkit, which
implements the 15 algorithms listed in Table~\ref{tab:algorithm}. The
toolkit is available for public usage and will be updated with
additional state-of-the-art algorithms proposed by the research
community.

There are three elementary baselines in Table~\ref{tab:algorithm}: a
constant function (identical prediction for all users and all
items), user average (constant prediction for each user-based on
their average ratings), and item average (constant prediction for
each item-based on their average ratings). The memory-based methods
listed in Table~\ref{tab:algorithm} are classical methods that
perform well and are often used in commercial settings. The methods
listed under the matrix factorization and others categories are more
recent state-of-the-art methods proposed in the research literature.

\begin{table}[!h]
\centering \footnotesize
\begin{tabular}{c|l} \hline
Category & Algorithms \tabularnewline \hline \hline

& Constant \\
Baseline & User Average \\
& Item Average \tabularnewline \hline

& User-based \cite{Su2009} \\
Memory & User-based w/ Default \cite{Breese1998} \\
-based& Item-based \cite{Sarwar2001} \\
& Item-based w/ Default \tabularnewline \hline

& Regularized SVD \cite{Paterek2007} \\
Matrix & NMF \cite{Lee2001} \\
Factorization & PMF \cite{Salakhutdinov2008a} \\
-based & Bayesian PMF \cite{Salakhutdinov2008b} \\
& Non-linear PMF \cite{Lawrence2009} \tabularnewline \hline

& Slope-One \cite{Lemire2005} \\
Others & NPCA \cite{Yu2009} \\
& Rank-based CF \cite{Sun2011} \tabularnewline \hline

\end{tabular}
\caption{List of Recommendation Algorithms used in Experiments}
\label{tab:algorithm}
\end{table}


In our experiments we used the Netflix dataset, a standard benchmark
in the CF literature that is larger and more recent than alternative
benchmarks. To facilitate measuring the dependency between
prediction accuracy and dataset size and density, we sorted the
rating matrix so that its rows and columns are listed in order of
descending density level. We then realized specific sparsity pattern
by selecting the top $k$ rows and $l$ columns and subsampling to
achieve the required sparsity.

\begin{figure*}
\centering{}
\includegraphics[width=17.5cm]{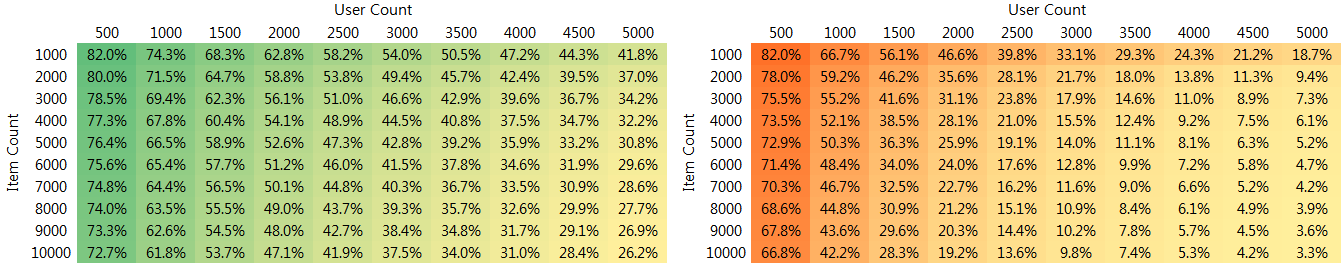}
\caption{Rating density (cumulative in the left panel and
non-cumulative in the right panel) in Netflix rating matrix, sorted by
descending density of rows and columns. See text for more detail.}
\label{fig:netflix_density}
\end{figure*}

Figure~\ref{fig:netflix_density} shows the density level of the
sorted rating matrix. For instance, the top right corner of the
sorted rating matrix containing the top 5,000 users and top 2,000
items has 52.6\% of density. In other words, there 47.4\% of the
ratings are missing. The density of the entire dataset is around
1\%. We subsample a prescribed level of density which will be used
for training as well as 20\% more for the purpose of testing. We
cross validate each experiment 10 times with different train-test
splits. The experiments were conducted on a dual Intel Xeon X5650
processor (6 Core, 12 Threads, 2.66GHz) with 96GB of main memory.

\subsection{Dependency on Data Size and Density}
\label{subsec:dependency}

We start by investigating the dependency of prediction accuracy on
the dataset size (number of users and number of items) and on the
rating density. Of particular interest, is the variability in that
dependency across different CF algorithms. This variability holds
the key to determining which CF algorithms should be used in a
specific situation. We start below by considering the univariate
dependency of prediction accuracy on each of these three quantities:
number of users, number of items, and density level. We then
conclude with an investigation of the multivariate dependency
between the prediction accuracy and these three variables.

\subsubsection{Dependency on User Count} \label{sec:userCount}

\begin{figure*}
    \centering{}
    \includegraphics[width=17.5cm]{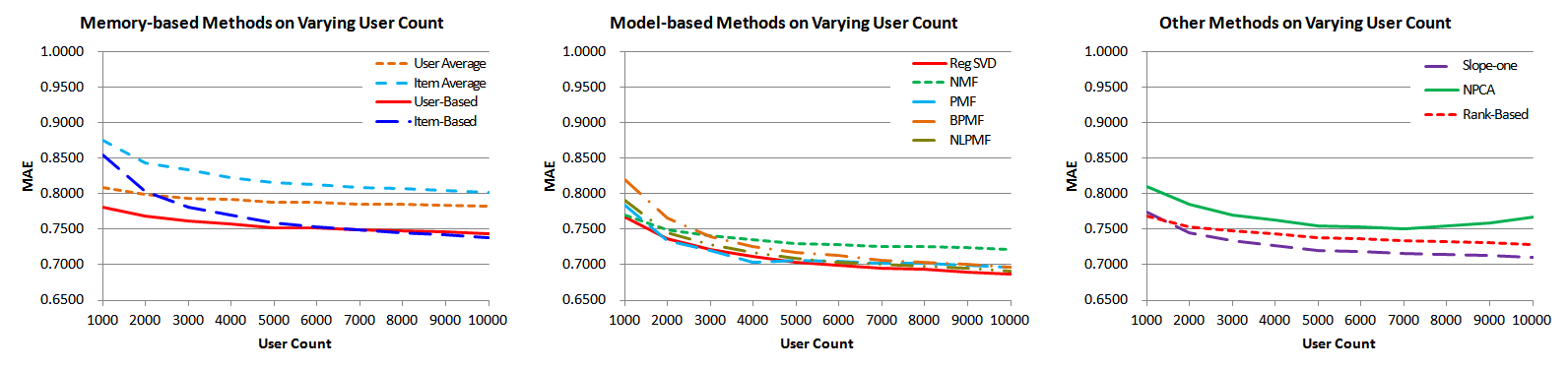}
    \includegraphics[width=17.5cm]{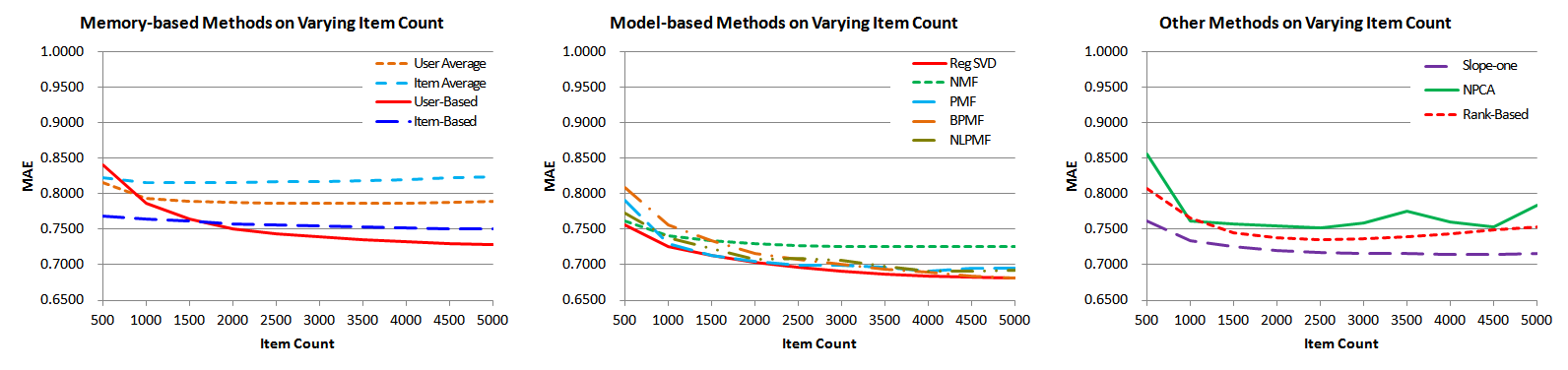}
    \includegraphics[width=17.5cm]{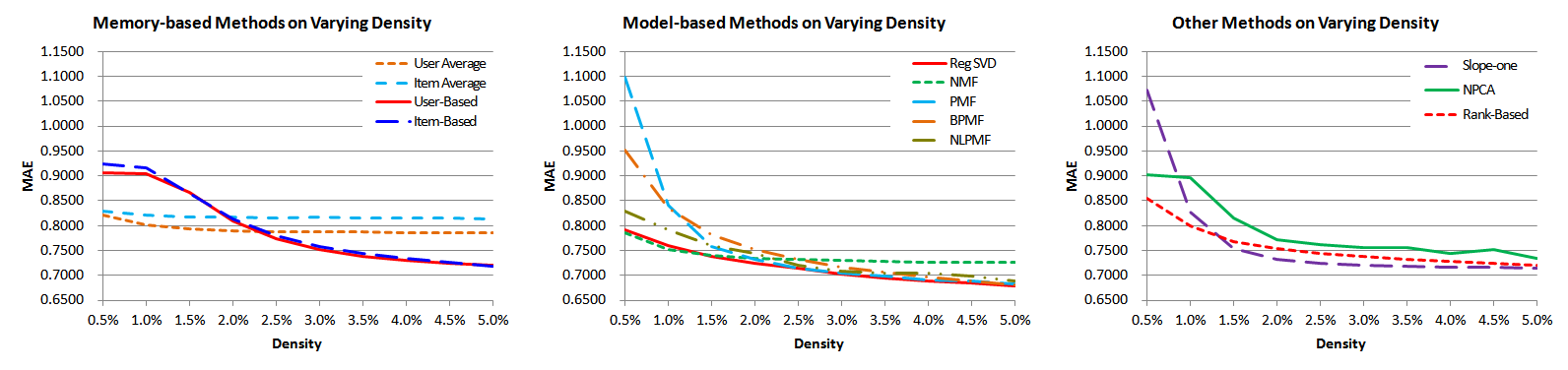}
    \caption{Prediction loss as a function of user count (top) item
      count (middle), density (bottom).}
    \label{fig:var}
\end{figure*}

Figure~\ref{fig:var} (top row) graphs the dependency of mean
absolute error (MAE) on the number of users with each of the three
panels focusing on CF methods in a specific category (memory-based
and baselines, model-based, and other). The item count and density
were fixed at 2,000 and 3\%, respectively. The RMSE evaluation
measure shows very similar trend.

We omitted the simplest baseline of constant prediction rule, since its
performance is much worse than the others. The default voting variants
of the user-based and item-based CF methods did not produce noticable
changes (we graphed the variant which worked the best).

To compare the way in which the different algorithms depend on the
user count quantity, we fitted a linear regression model to the curves
in Figure~\ref{fig:var} (top row).  The slope $m$ and intercept $b$
regression coefficients appear in Table~\ref{tab:dependency_reg}. The
intercept $b$ indicates the algorithm's expected MAE loss when the
number of users approach 0. The slope $m$ indicates the rate of decay
of the MAE.

Looking at Figure~\ref{fig:var} and Table~\ref{tab:dependency_reg}, we
make the following observations.

\begin{enumerate}
\item Matrix factorization methods in general show better performance
  when the number of users gets sufficiently large ($>3,000$).
\item Overall, the best performing algorithm is regularized SVD.
\item When the number of users is sufficiently small there is very
  little difference between matrix factorization methods and the
  simpler neighborhood-based methods.
\item Item average, item-based, regularized SVD, PMF, BPMF, and NLPMF
  tend to be the most sensitive to variation in user count.
\item Constant baseline, user average, user-based, NMF, NPCA, and
  rank-based are relatively insensitive to the number of users.
\item There is stark difference in sensitivity between the two popular
neighborhood-based methods: user-based CF and item-based CF.
\item User-based CF is extremely effective for low user count but has an
  almost constant dependency on the user count. Item-based CF
  performs considerably worse at first, but outperforms all other
  memory-based methods for larger user count.
\end{enumerate}

\begin{table*}[!htb]
\centering \footnotesize
\begin{tabular}{l|r|r|r|r|r|r} \hline
Algorithm & \multicolumn{2}{|c|}{User Count} & \multicolumn{2}{|c|}{Item Count} & \multicolumn{2}{|c}{Density} \\

\cline{2-7}

& $10^{-6}m$ & $b$ & $10^{-6}m$ & $b$ & $m$ & $b$\tabularnewline
\hline \hline


Constant                    &  -1.7636 & 0.9187 & -13.8097 & 0.9501 & -0.0691 & 0.9085\\
User Average                &  -2.5782 & 0.8048 &  -3.5818 & 0.8006 & -0.6010 & 0.8094\\
Item Average                &  -6.6879 & 0.8593 &  +1.1927 & 0.8155 & -0.2337 & 0.8241\\
User-based                  &  -3.9206 & 0.7798 & -18.1648 & 0.8067 & -4.7816 & 0.9269\\
User-based (Default values) &  -3.6376 & 0.7760 & -19.3139 & 0.8081 & -4.7081 & 0.9228\\
Item-based                  & -10.0739 & 0.8244 &  -3.2230 & 0.7656 & -4.7104 & 0.9255\\
Item-based (Default values) & -10.4424 & 0.8271 &  -3.6473 & 0.7670 & -4.9147 & 0.9332\\
Slope-one                   &  -5.6624 & 0.7586 &  -7.5467 & 0.7443 & -5.1465 & 0.9112\\
Regularized SVD             &  -7.6176 & 0.7526 & -14.1455 & 0.7407 & -2.2964 & 0.7814\\
NMF                         &  -4.4170 & 0.7594 &  -5.7830 & 0.7481 & -0.9792 & 0.7652\\
PMF                         &  -6.9000 & 0.7531 & -14.7345 & 0.7529 & -6.3705 & 0.9364\\
Bayesian PMF                & -11.0558 & 0.7895 & -23.7406 & 0.7824 & -4.9316 & 0.8905\\
Non-linear PMF              &  -8.8012 & 0.7664 & -14.7588 & 0.7532 & -2.8411 & 0.8135\\
NPCA                        &  -4.1497 & 0.7898 &  -7.2994 & 0.7910 & -3.5036 & 0.8850\\
Rank-based CF               &  -3.8024 & 0.7627 &  -7.3261 & 0.7715 & -2.4686 & 0.8246

\tabularnewline \hline

\end{tabular}
\caption{Regression coefficients $y = mx + b$ for the curves in
  Figure~\ref{fig:var}. The variable $x$ represents user count, item
  count, or density, and the variable $y$ represents the MAE
  prediction loss.}
\label{tab:dependency_reg}
\end{table*}

\subsubsection{Dependency on Item Count} \label{sec:itemCount}

In analogy with Section~\ref{sec:userCount} we investigate here the
dependency of the prediction loss on the number of items, fixing the
user count at and density at 5,000 and  3\%, respectively.  Figure~\ref{fig:var} (middle row) shows the MAE as a function of the
number of items for three different categories of CF
algorithms. Table~\ref{tab:dependency_reg} shows the regression
coefficients (see description in Section~\ref{sec:userCount}).

Looking at Figure~\ref{fig:var} and Table~\ref{tab:dependency_reg}, we
make the following observations that are largely in agreement with the
observations in Section~\ref{sec:userCount}.

\begin{enumerate}
\item Matrix factorization methods in general show better performance
  when the number of items gets sufficiently large ($>1,000$).
\item Overall, the best performing algorithm is regularized SVD.
\item When the number of users is sufficiently small there is very
  little difference between matrix factorization methods and the
  simpler neighborhood-based methods.
\item User-based, regularized SVD, PMF, BPMF, and NLPMF tend to be the most sensitive to variation in item count.
\item Item average, item-based, and NMF tend to be less
  sensitive to variation in item count.
\item There is stark difference in sensitivity between the two popular
neighborhood-based methods: user-based CF and item-based CF.
\item Item-based CF is extremely effective for low item count but has an
  almost constant dependency on the item count. User-based CF
  performs considerably worse at first, but outperforms all other
  memory-based methods significantly for larger user count.
\item Combined with the observations in Section~\ref{sec:userCount},
  we conclude that slope-one, NMF, and NPCA are insensitive to
  variations in both user and item count.  PMF and BPMF are relatively
  sensitive to variations in both user and item count.
\end{enumerate}

\subsubsection{Dependency on Density} \label{sec:density}

Figure~\ref{fig:var} (bottom row) graphs the dependency of the MAE
loss on the rating density, fixing user and item count at 5,000 and
2,000. As in Section~\ref{sec:userCount},
Table~\ref{tab:dependency_reg} displays regression coefficients
corresponding to performance at near 0 density and sensitivity of
the MAE function to the density level.

Looking Figure~\ref{fig:var} (bottom row) and
Table~\ref{tab:dependency_reg} we make the following observations.

\begin{enumerate}
\item The simple baselines (user average and item average) work
  remarkably well for low density.
\item The best performing algorithms seem to be regularized SVD.
\item User-based CF and item-based CF show a remarkable similar
  dependency on density level. This is in stark contrast to their
  different dependencies on the user and item count.
\item As the density level increases the differences in prediction
  accuracy of the different algorithm shrink.
\item User-based, item-based, slope-one, PMF, and BPMF are largely
dependent on density.
\item The three baselines and NMF are relatively independent from density.
\item The performance of slope-one and PMF degrade significantly at
  low densities, performing worse than the weakest
  baselines. Nevertheless both algorithms feature outstanding
  performance at high densities.
\end{enumerate}

\subsubsection{Mutlivariate Dependencies between Prediction Loss, User Count, Item Count, and Density}

\begin{figure}
    \centering{}
    \includegraphics[width=17cm]{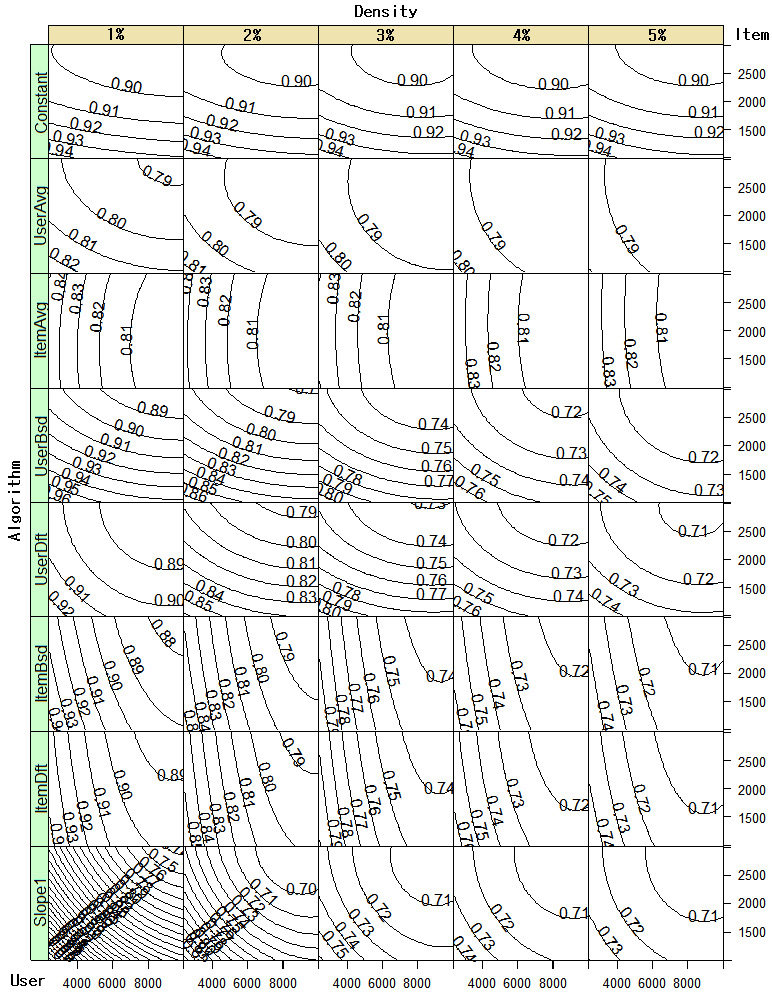}
    \caption{MAE Contours for simple method (Lower values mean better performance.)}
    \label{fig:mae1}
\end{figure}

\begin{figure}
    \centering{}
    \includegraphics[width=17cm]{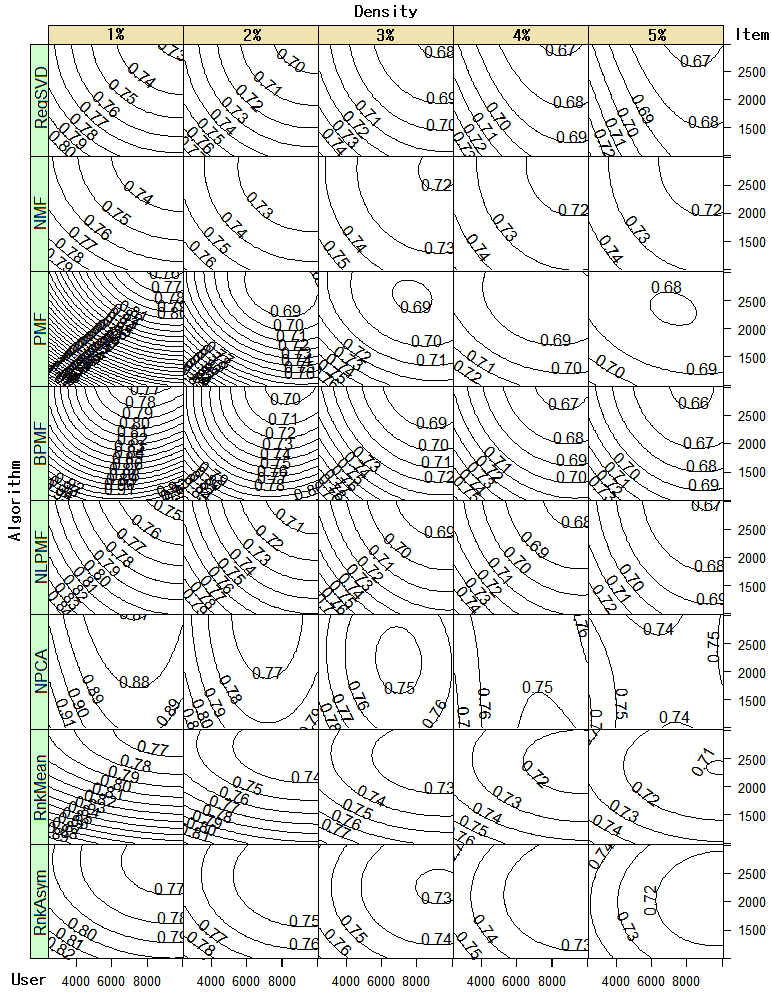}
    \caption{MAE Contours for advanced method (Lower values mean better performance.)}
    \label{fig:mae2}
\end{figure}

The univariate dependencies examined previously show important
trends but are limited since they examine variability of one
quantity while fixing the other quantities to arbitrary values. We
now turn to examine the dependency between prediction loss and the
following variables: user count, item count, and density. We do so
by graphing the MAE as a function of user count, item count, and
density (Figure~\ref{fig:mae1}--\ref{fig:mae2}) and by fitting
multivariate regression models to the dependency of MAE on user
count, item count, and density
(Table~\ref{tab:dependency_combined}).

Figure~\ref{fig:mae1}--\ref{fig:mae2} shows the equal height
contours of the MAE as a function of user count ($x$ axis) and item
count ($y$ axis) for multiple density levels (horizontal panels) and
for multiple CF algorithms (vertical panels). Note that all contour
plots share the same $x$ and $y$ axes scales and so are directly
comparable. Intervals between different contour lines represent a
difference of 0.01 in MAE and so more contour lines represent higher
dependency on the $x$ and $y$ axis. Analogous RMSE graphs show
similar trend to these MAE graphs.

Table~\ref{tab:dependency_combined} displays the regression
coefficients $m_u, m_i, m_d$ corresponding to the linear model
\begin{equation}
y = m_u x_u + m_i x_i + m_d x_d + b \label{eq:regression_combined}
\end{equation}
where $x_u$, $x_i$, and $x_d$ indicate user count, item count, and
density, $b$ is the constant term, and $y$ is the MAE.

Based on Figure~\ref{fig:mae1}--\ref{fig:mae2} and
Table~\ref{tab:dependency_combined} we make the following
observations.

\begin{enumerate}
\item The univariate relationships discovered in the previous section for fixed values of the remaining two variables do not necessarily hold in general. For example, the conclusion that PMF is relatively sensitive to user count and item count at 1\% sparsity level and 3\% sparsity level is not longer valid for 5\% density levels. It is thus important to conduct a multivariate, rather than univariate, analysis of the dependency of the prediction loss on the problem parameters.
\item The shape of the MAE contour curves vary from algorithm to algorithm. For example, the contour curves of constant, user average, user-based, and user-based with default values are horizontal, implying that these algorithms depend largely on the number of items, regardless of user count. On the other hand, item average, item-based, and item-based with default values show vertical contour lines, showing a high dependency on the user count.
\item Higher dependency on user count and item count is correlated with high dependency on density.
\item The last column of Table~\ref{tab:dependency_combined} summarizes dependency trends based on the absolute value of the regression coefficients and their rank. Generally speaking, baselines are relatively insensitive, memory-based methods are dependent on one variable (opposite to their names), and matrix factorization methods are highly dependent on both dataset size and density.
\end{enumerate}

\begin{table*}[!htb]
\centering \scriptsize
\begin{tabular}{l|r|r|r|r|l} \hline

Algorithm & $m_u$ & $m_i$ & $m_d$ & $b$ & Summary\tabularnewline
\hline \hline

Constant                & (15) -0.0115 &  (8) -0.0577 & (15) -0.0002 & 0.9600 & Weekly dependent on all variables.\\
User Average            & (13) -0.0185 & (14) -0.0164 & (13) -0.0188 & 0.8265 & Weekly dependent on all variables.\\
Item Average            &  (8) -0.0488 & (15) +0.0003 & (14) -0.0078 & 0.8530 & Weekly dependent on all variables.\\
User-based              & (11) -0.0282 &  (3) -0.0704 &  (1) -0.2310 & 0.9953 & Weekly dependent on user count.\\
User-based (w/Default)  & (12) -0.0260 &  (7) -0.0598 &  (5) -0.2185 & 0.9745 & Weekly dependent on user count.\\
Item-based              &  (3) -0.0630 & (12) -0.0172 &  (4) -0.2201 & 0.9688 & Weekly dependent on item count.\\
Item-based (w/Default)  &  (2) -0.0632 & (13) -0.0167 &  (2) -0.2286 & 0.9751 & Weekly dependent on item count.\\
Slope-one               &  (1) -0.0746 &  (4) -0.0702 &  (8) -0.1421 & 0.9291 & Strongly dependent on all variables.\\
Regularized SVD         &  (7) -0.0513 &  (9) -0.0507 & (11) -0.0910 & 0.8371 & Strongly dependent on dataset size.\\
NMF                     &  (9) -0.0317 & (10) -0.0283 & (12) -0.0341 & 0.7971 & Weekly dependent on all variables.\\
PMF                     &  (5) -0.0620 &  (2) -0.1113 &  (3) -0.2269 & 0.9980 & Strongly dependent on all variables.\\
Bayesian PMF (BPMF)     &  (4) -0.0628 &  (1) -0.1126 &  (6) -0.1999 & 0.9817 & Strongly dependent on all variables.\\
Non-linear PMF (NLPMF)  &  (6) -0.0611 &  (6) -0.0599 &  (9) -0.1165 & 0.8786 & Strongly dependent on dataset size.\\
NPCA                    & (14) -0.0184 & (11) -0.0213 &  (7) -0.1577 & 0.9103 & Weekly dependent on dataset size.\\
Rank-based CF           & (10) -0.0295 &  (5) -0.0687 & (10) -0.1065 & 0.8302 & Strongly dependent on item count.

\tabularnewline \hline

\end{tabular}
\caption{Regression coefficients for the model $y = m_u z_u + m_i z_i + m_d z_d + b$ \eqref{eq:regression_combined}) where $y$
is MAE, $z_u$, $z_i$, and $z_d$ are inputs from
$x_u$, $x_i$, and $x_d$, normalized to achieve similar scales:  $z_u = x_u / 10,000$, $z_i = x_i / 3,000$, and $z_d = x_d /
0.05$. The rank on each variable is indicated in parenthesis with rank
1 showing highest dependency.)} \label{tab:dependency_combined}
\end{table*}

\subsection{Accuracy Comparison}
\label{subsec:accuracy}

\begin{figure*}
    \centering{}
    \includegraphics[width=17.5cm]{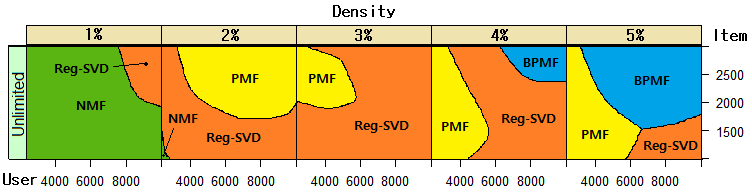}
    \caption{Best-performing algorithms in MAE for given user count, item count, and density.}
    \label{fig:best_unlimited}
\end{figure*}

Figure~\ref{fig:best_unlimited} shows the best performing algorithm
(in terms of MAE) as a function of user count, item count,
and density. We make the following conclusions.

\begin{enumerate}
\item The identity of the best performing algorithm varies is non-linearly dependent on user count, item count, and density.
\item NMF is dominant low density cases while BPMF works well for high density cases (especially for high item and user count).
\item Regularized SVD and PMF perform well for density levels 2\%-4\%.

\end{enumerate}

Analogous RMSE graphs show similar trends with regularized SVD outperforming other algorithms in most regions.

\subsection{Asymmetric and Rank-based Metrics}
\label{subsec:metrics}

We consider here the effect of replacing the MAE or RMSE with other
loss functions, specifically with asymmetric loss and with
rank-based loss.

Asymmetric loss is motivated with the fact that recommending an
undesirable item is worse than avoiding recommending a desirable
item. In other words, the loss function $L(a,b)$, measuring the
effect of predicting rating $b$ when the true rating is $a$ is an
asymmetric function. Specifically, we consider the loss function
$L(a,b)$ defined by the following matrix (rows and columns express
number of stars on a 1-5 scale)
\begin{equation}
L(\cdot,\cdot)=\begin{pmatrix}
0 & 0 & 0 & 7.5 & 10 \\
0 & 0 & 0 & 4 & 6 \\
0 & 0 & 0 & 1.5 & 3 \\
3 & 2 & 1 & 0 & 0 \\
4 & 3 & 2 & 0 & 0
\end{pmatrix}.
\label{eq:asym}
\end{equation}
This loss function represents two beliefs: 1) Difference among items
to be recommended is not important. Assuming that we issue
recommendations with rating 4 or 5, no loss is given between the
two. In the same way, we do not penalize error among items which
will not be recommended. 2) We give severer penalty for recommending
bad items than for missing potentially preferable items. For the
latter case, the loss is the exact difference between the prediction
and ground truth. For the former case, however, we give higher
penalty. For example, penalty is 10 for predicting worst item with
true score 1 as score 5, higher than 4 for the opposite way of
prediction. In many practical cases involving recommendation
systems, asymmetric loss functions provide a more realistic loss
function than symmetric loss functions such as MAE or RMSE.

Rank-based loss function are based on evaluating a ranked list of
recommended items, presented to a user. The evaluation of the list
gives higher importance to good recommendations at the top of the
list, than at the bottom of the list. One specific formula, called
half life utility (HLU)~\cite{Breese1998,Huang2007} assumes an
exponential decay in the list position. Formally, the utility
function associated with a user $u$
\begin{equation}
R_u = \sum_{i=1}^{N} \frac{max(r_{u,i} - d,
0)}{2^{(i-1)/(\alpha-1)}} \label{eq:rank}
\end{equation}
where $N$ is the number of recommended items (length of the list),
$r_{u,i}$ is the rating of user $u$ for item $i$ in the list, and
$d$ and $\alpha$ are constants, set to $d = 3$, and $\alpha = 5$ (we
assume $N=10$). The final utility function is $R_u$ divided by the
maximum possible utility for the user, average over all test users
\cite{Breese1998}. Alternative rank-based evaluations are based on
NDCG~\cite{Jarvelin2002}, and Kendall's Tau, and Spearman's Rank
Correlation Coefficient~\cite{Marden1996}.

\subsubsection{Asymmetric Loss}

\begin{figure}
    \centering{}
    \includegraphics[width=17cm]{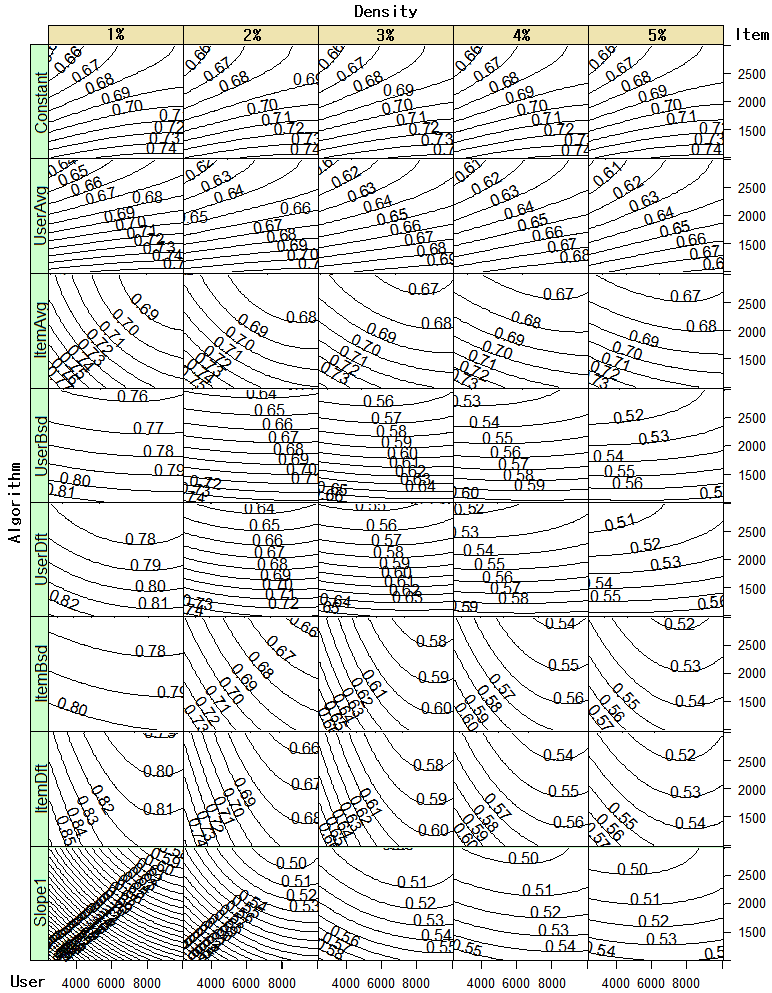}
    \caption{Asymmetric Loss Contours for simple methods (Lower values mean better performance.)}
    \label{fig:asym1}
\end{figure}

\begin{figure}
    \centering{}
    \includegraphics[width=17cm]{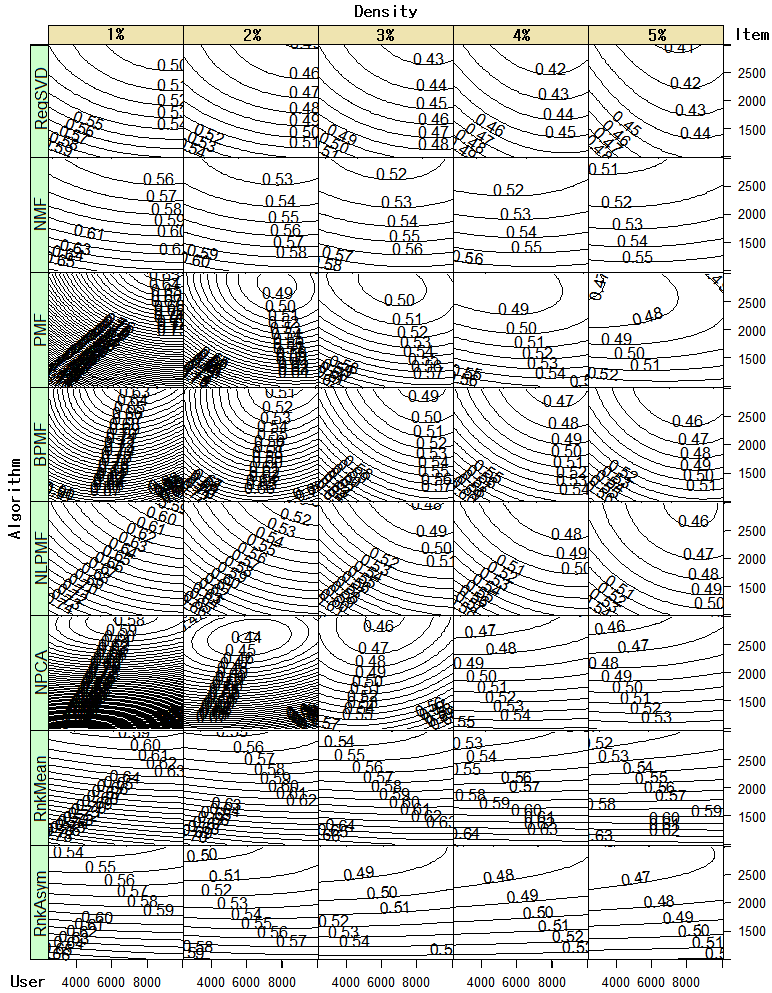}
    \caption{Asymmetric Loss Contours for simple methods (Lower values mean better performance.)}
    \label{fig:asym2}
\end{figure}

Figure~\ref{fig:asym1}--\ref{fig:asym2} shows equal level contour
plots of the asymmetric loss function \eqref{eq:asym}, as a function
of user count, item count, and density level. We make the following
observations.

\begin{enumerate}
\item The shape and density pattern of the contour lines differ from the shape of the contour lines in the case of the MAE.
\item In general, regularized SVD outperforms all other algorithms.
Other matrix factorization methods (PMF, BPMF, and NLPMF) perform
relatively well for dense data. With sparse data, NMF performs well.
\end{enumerate}

\subsubsection{Rank-based Evaluation Measures}

\begin{figure}
    \centering{}
    \includegraphics[width=17cm]{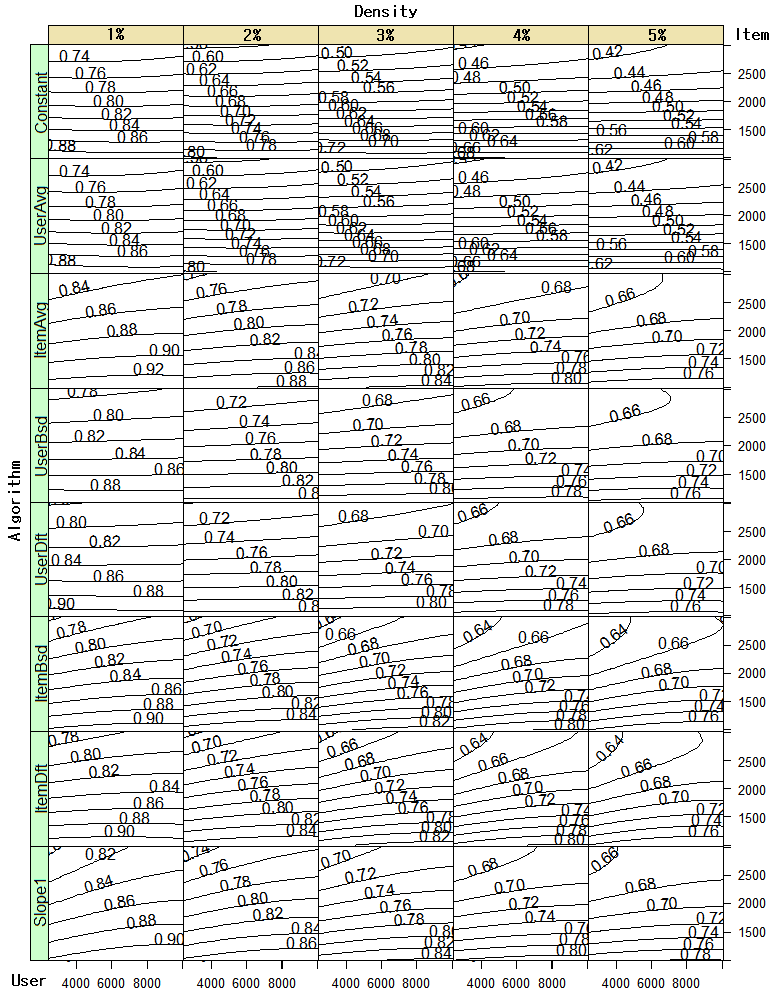}
    \caption{Half-Life Utility Contours for simple method (Higher values mean better performance.)}
    \label{fig:hlu1}
\end{figure}

\begin{figure}
    \centering{}
    \includegraphics[width=17cm]{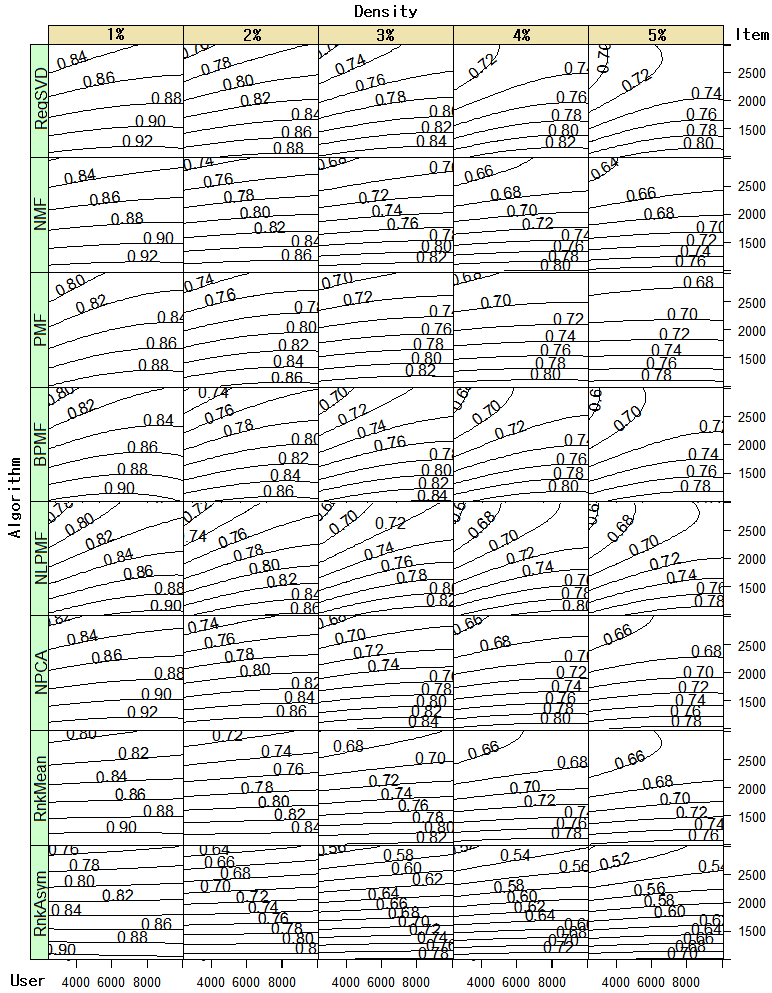}
    \caption{Half-Life Utility Contours for advanced method (Higher values mean better performance.)}
    \label{fig:hlu2}
\end{figure}

\begin{figure*}
    \centering{}
    \includegraphics[width=17.5cm]{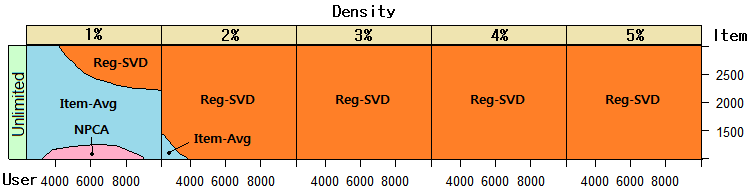}
    \caption{Best-performing algorithms in HLU for given user count, item count, and density.}
    \label{fig:best_hlu}
\end{figure*}

Figure~\ref{fig:hlu1}--\ref{fig:hlu2} show equal level contours of
the HLU function \eqref{eq:rank}. Figure~\ref{fig:best_hlu} shows
the best performing algorithm for different user count, item count,
and density. We make the following observations.

\begin{enumerate}
\item The contour lines are generally horizontal, indicating that performance under HLU depend largely on the number of items and is less affected by the number of users.
\item The HLU score is highly sensitive to the dataset density.
\item  Regularized SVD outperforms other methods (see Figure~\ref{fig:best_hlu}) in most settings. The simple baseline item average is best for small and sparse datasets. A similar comment can be made regarding NPCA. NMF and slope-one perform well for sparse data, though they lag somewhat the previous mentioned algorithms.
\end{enumerate}

Other rank-based loss functions based on NDCG, Kendall's Tau, and
Spearman show similar trends.

\subsection{Computational Considerations}
\label{subsec:computation}

\begin{figure*}
    \centering{}
    \includegraphics[width=17.5cm]{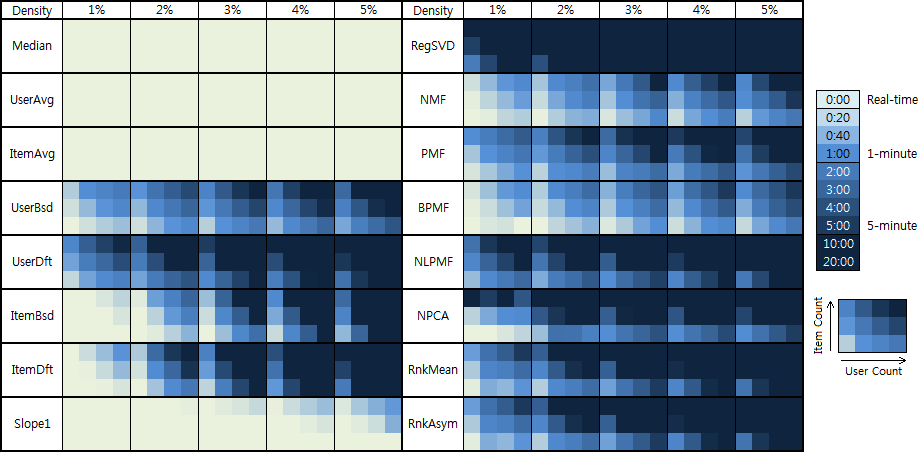}
    \caption{Computation time (Train time + Test time) for each algorithm.
    Legend on the right indicates relation between color scheme and computation time.
    Time constraints (5 minutes and 1 minutes) used in this article
    are marked as well. User count increases from left to right, and item count increases
    from bottom to top in each cell. (Same way to Figure~\ref{fig:mae1}--\ref{fig:mae2}.)}
    \label{fig:time}
\end{figure*}

As Figure~\ref{fig:time} shows, the computation time varies significantly between different algorithms. It is therefore important to consider computational issues when deciding on the appropriate CF algorithm. We consider three distinct scenarios, listed below.\\

\begin{itemize}
\setlength{\itemsep}{0cm}
\setlength{\parskip}{0cm}

\item \textbf{Unlimited Time Resources:} We assume in this case that we can afford arbitrarily long computation time. This scenario is realistic in some cases involving static training set, making offline computation feasible.

\item \textbf{Constrained Time Resource:} We assume in this cases
some mild constraints on the computation time. This scenario is
realistic in cases  where the training set is periodically updated,
necessitating periodic re-training with updated data. We assume here
that the training phase should tale within an hour or so. Since
practical datasets like Netflix full set are much bigger than the
subsampled one in our experiments, we use much shorter time limit: 5
minutes and 1 minute.

\item \textbf{Real-time Applications:} We assume in this case that severe constraints on the computation time. This scenario is realistic in cases where the training set changes continuously. We assume here that the training phase should not exceed several seconds.\\
\end{itemize}

Figure~\ref{fig:best_unlimited} and~\ref{fig:time} show the best
performing CF algorithm (in terms of MAE) in several different time
constraint cases, as a function of the user count, item count and
density.

\begin{figure*}
    \centering{}
    \includegraphics[width=17.5cm]{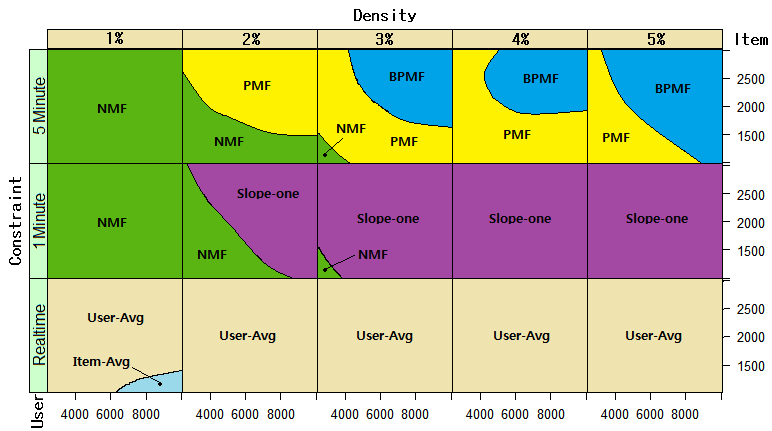}
    \caption{Best-performing algorithms with varied constraints.}
    \label{fig:best_constrained}
\end{figure*}

We make the following observations.

\begin{enumerate}

\item When there are no computation constraints, the conclusions from the previous sections apply. Specifically, NMF performs best for sparse dataset, BPMF performs best for dense dataset, and regularized
SVD and PMF perform the best otherwise (PMF works well with smaller user count while Regularized SVD works well smaller item counts).

\item When the time constraint is 5 minutes, Regularized SVD, NLPMF,
NPCA, and Rank-based CF (the ones colored darkly in
Figure~\ref{fig:time}) are not considered. In this setting, NMF
works best for sparse data, BPMF works best for dense and large data
and PMF works best otherwise.

\item When the time constraint is 1 minutes, PMF and BPMF are additionally excluded from
consideration. Slope-one works best in most cases, except for the
sparsest data where NMF works best.

\item In cases requiring real-time computation, the user average is the best algorithm, except for a small region where item average is preferred.
\end{enumerate}





\section{Discussion}
\label{sec:conclusion}

In addition to the conclusions stated in Section~\ref{sec:exp}, we have identified seven groups of CF methods, where CF methods in the same group share certain experimental properties:

\begin{itemize}
  \setlength{\itemsep}{0cm}
  \setlength{\parskip}{0cm}

  \item Baselines: Constant, User Average, Item Average
  \item Memory-based methods: User-based, Item-based (with and
  without default values)
  \item Matrix-Factorization I: Regularized SVD, PMF, BPMF, NLPMF
  \item Matrix-Factorization II: NMF
  \item Others (Individually): Slope-one, NPCA, Rank-based CF.
\end{itemize}

Table~\ref{tab:summary} displays for each of these groups the dependency, accuracy, computational cost, and pros and cons.

\begin{sidewaystable*}
\centering \footnotesize
\begin{tabular}{c|c|c|c|c|c|c|c|c} \hline \hline
\multicolumn{2}{c|}{Category} & Baselines & Memory-based &
\multicolumn{2}{|c|}{Matrix-factorization} &
\multicolumn{3}{|c}{Others} \tabularnewline \hline

\multicolumn{2}{c|}{Algorithms} & Constant, & User-based, & Reg-SVD, & NMF & Slope-one & NPCA & Rank-based \\
\multicolumn{2}{c|}{} & User Average, & Item-based, & PMF, BPMF, & & & \\
\multicolumn{2}{c|}{} & Item Average & w/default & NLPMF & & &
\tabularnewline \hline \hline

Dependency & Size & \textbf{Very low} & Low & High & Low & High &
\textbf{Very low} & Fair \tabularnewline \cline{2-9}

& Density & \textbf{Very low} & High & Very high & Low & Very high &
High & Fair \tabularnewline \hline

Accuracy & Dense & Very poor & Good & \textbf{Very good} & Good &
Good & Poor & Fair \tabularnewline \cline{2-9}

& Sparse & Poor & Very poor & Fair & \textbf{Very good} & Poor &
Poor & Fair \tabularnewline \hline

\multicolumn{2}{c|}{Asymmetric accuracy} & Poor & Fair &
\textbf{Very good} & Good & Good & \textbf{Very good} & \textbf{Very
good} \tabularnewline \hline

\multicolumn{2}{c|}{HLU/NDCG} & Very poor & Fair & \textbf{Very
good} & Fair & Good & Fair & Fair \tabularnewline \hline

\multicolumn{2}{c|}{Kendall's Tau/Spearman} & Very poor & Fair &
Good & Fair & Fair & Fair & Fair \tabularnewline \hline

Computation & Train & \textbf{No} & \textbf{No} & Slow & Fast &
\textbf{Very fast} & Slow & No \tabularnewline \cline{2-9}

& Test & \textbf{Very fast} & Very slow & Fair & Fair & Fast & Slow
& Very slow \tabularnewline \hline

\multicolumn{2}{c|}{Memory consumption} & \textbf{Low} & High & High
& High & \textbf{Low} & Very high & High \tabularnewline \hline

\multicolumn{2}{c|}{Adjustable parameters} & \textbf{No} & Few &
Many & Many & \textbf{No} & Few & Few \tabularnewline \hline

\multicolumn{2}{c|}{Overall Merits}
 & Computation
 & Do not need
 & Perform best
 & Perform best
 & Perform well
 & Perform well
 & Perform well
\tabularnewline

\multicolumn{2}{c|}{}
 & takes
 & to train.
 & with high-
 & with sparse
 & in spite of
 & when using
 & when using
\tabularnewline

\multicolumn{2}{c|}{}
 & little time.
 &
 & density data.
 & data. Train
 & short time.
 & asymmetric
 & asymmetric
\tabularnewline

\multicolumn{2}{c|}{}
 &
 &
 &
 & is fast.
 &
 & measures.
 & measures.
\tabularnewline \hline

\multicolumn{2}{c|}{Overall Demerits}
 & Accuracy is
 & Testing takes
 & Many parameters
 & Many
 & Perform poorly
 & Uses extremely
 & Computation
\tabularnewline

\multicolumn{2}{c|}{}
 & very poor.
 & very long time.
 & should be
 & parameters
 & without
 & large memory
 & takes too
\tabularnewline

\multicolumn{2}{c|}{}
 &
 & Uses lots
 & adjusted.
 & should be
 & large/dense
 & during
 & long time.
\tabularnewline

\multicolumn{2}{c|}{}
 &
 & of memory.
 & Computation
 & adjusted.
 & dataset.
 & training.
 &
\tabularnewline

\multicolumn{2}{c|}{}
 &
 &
 & takes long.
 &
 &
 &
 &
\tabularnewline \hline \hline

\end{tabular}
\caption{Summary of Pros and Cons for Recommendation Algorithms}
\label{tab:summary}
\end{sidewaystable*}

We repeat below some of the major conclusions. See
Section~\ref{sec:exp} for more details and additional conclusions.

\begin{itemize}
  \setlength{\itemsep}{0cm}
  \setlength{\parskip}{0cm}

\item Matrix-Factorization-based methods generally have the highest accuracy. Specifically,
regularized SVD, PMF and its variations perform best as far as MAE
and RMSE, except in very sparse situations, where NMF performs the
best. Matrix-factorization methods perform well also in terms of the
asymmetric cost and rank-based evaluation measures. NPCA and
rank-based CF work well in these cases as well. The Slope-one method
performs well and is computationally efficient. Memory-based
methods, however, do not have special merit other than simplicity.

\item All algorithms vary in their accuracy, based on the user count, item count, and density. The strength and nature of the dependency, however, varies from algorithm to algorithm and bivariate relationships change when different values are assigned to the third variable. In general cases, high dependence on the user count and item count is correlated with high dependency on density, which appeared to be the more influential
factor.

\item There is trade-off between better accuracy and other
factors such as low variance in accuracy, computational efficiency,
memory consumption, and a smaller number of adjustable parameters.
That is, the more accurate algorithms tend to depend highly on
dataset size and density, to have higher variance in accuracy, to be
less computationally efficient, and to have more adjustable
parameters. A careful examination of the experimental results can
help resolve this tradeoff in a manner that is specific to the
situation at hand. For example, when computational efficiency is
less important, Matrix-Factorization methods are the most
appropriate, and when computational efficiency is important,
slope-one could be a better choice.
\end{itemize}



This experimental study, accompanied by an open source software that allows reproducing our experiments, sheds light on how CF algorithms compare to each other, and on their dependency on the problem parameters. The conclusions described above should help practitioners, implementing recommendation systems, and researchers examining novel state-of-the-art CF methods.

\bibliographystyle{abbrv}
\bibliography{../../share/externalPapers,../../share/groupPapers}

\end{document}